\documentclass[twocolumn,prl,showpacs,floatfix,amsfonts,superscriptaddress]{revtex4}

\usepackage{graphicx}

\begin{document}
\title{Probing the $\gamma$-$\alpha$ Transition in Bulk Ce under Pressure: \\
A Direct Investigation by Resonant Inelastic X-ray Scattering}
\author{J.-P. Rueff}
\affiliation{Synchrotron SOLEIL, L'Orme des Merisiers, Saint-Aubin, BP~48, 91192 Gif-sur-Yvette Cedex, France} 
\affiliation{Laboratoire de Chimie Physique--Mati\`ere et Rayonnement (UMR~7614), Universit\'e Pierre et Marie Curie, 11 rue Pierre et Marie Curie, 75231 Paris Cedex~05, France}
\author{J.-P. Iti\'e}
\affiliation{Synchrotron SOLEIL, L'Orme des Merisiers, Saint-Aubin, BP~48, 91192 Gif-sur-Yvette Cedex, France} 
\author{M. Taguchi}
\affiliation{Soft X-ray spectroscopy Lab, RIKEN/SPring-8, 1-1-1, Sayo, Sayo, Hyogo 679-5148, Japan}
\author{C. F. Hague}
\affiliation{Laboratoire de Chimie Physique--Mati\`ere et Rayonnement (UMR~7614), Universit\'e Pierre et Marie Curie, 11 rue Pierre et Marie Curie, 75231 Paris Cedex~05, France}
\author{J.-M. Mariot}
\affiliation{Laboratoire de Chimie Physique--Mati\`ere et Rayonnement (UMR~7614), Universit\'e Pierre et Marie Curie, 11 rue Pierre et Marie Curie, 75231 Paris Cedex~05, France}
\author{R. Delaunay}
\affiliation{Laboratoire de Chimie Physique--Mati\`ere et Rayonnement (UMR~7614), Universit\'e Pierre et Marie Curie, 11 rue Pierre et Marie Curie, 75231 Paris Cedex~05, France}
\author{J.-P. Kappler}
\affiliation{IPCMS (UMR~7504), 23 rue du L{\oe}ss, BP~43, 67034 Strasbourg Cedex, France}
\author{N. Jaouen}
\affiliation{ESRF, 6 rue Jules Horowitz, BP~220, 38043 Grenoble Cedex, France}
\affiliation{Synchrotron SOLEIL, L'Orme des Merisiers, Saint-Aubin, BP~48, 91192 Gif-sur-Yvette Cedex, France}

\date{\today}

\begin{abstract}
We report on the most complete investigation to date of the $4f$-electron properties at the $\gamma$-$\alpha$ transition in elemental Ce by resonant inelastic x-ray scattering (RIXS). The Ce $2p3d$-RIXS spectra were measured directly in the bulk material as a function of pressure through the transition. The spectra were simulated within the Anderson impurity model. The occupation number $n_f$ was derived from the calculations in both $\gamma$- and $\alpha$-phases in the ground state along with the $f$ double-occupancy. We find that the electronic structure changes result mainly from band formation of $4f$ electrons which concurs with reduced electron correlation and increased Kondo screening at high pressure.
\end{abstract}
\pacs{78.70.En, 71.27.+a, 71.20.Eh}
\maketitle

The $\gamma$-$\alpha$\ transition in Ce is archetypical of localization-delocalization phenomenon encountered in $f$-electron systems. This isostructural phase transition accompanied by a large volume contraction that ends in a tri-critical point is a manifestation of subtle interactions between $f$-levels self-consistently embedded in a sea of conduction electrons. The Ce anomalous behavior at the transition is commonly described by one or the other of two possible scenarios involving either a Mott transition~\cite{Johansson1974} or Kondo hybridization~\cite{Allen1982,Lavagna1982}.  Recent theoretical approaches seem to point to a somewhat intermediate behavior at finite temperature in either phase, with electron correlations and screening as mandatory ingredients~\cite{Medici2005,McMahan2003}. More precisely, both phases can be described by strongly correlated $4f$ electrons, but differ in their degree of localization: the occupation number $n_f$ is almost equal to unity in the $\gamma$-phase and is reduced in the mixed-valent regime in the $\alpha$-phase. As stated in Ref.~\onlinecite{McMahan2003}, important information about the $\gamma$-$\alpha$\ transition and more particularly the effects of electron correlation in Ce, is contained not only in $n_f$, but also in the probability of double occupancy of the $f$ sites. $4f$ electron states are clearly identifiable in spectroscopic data  obtained by x-ray photoelectron (XPS) or x-ray absorption (XAS) spectroscopies. They are thus the two probes that have contributed most to unraveling the electronic properties of Ce in the past, and proof of a mixed-valent behavior has been accumulated unambiguously from such experiments over the years. Well separated features, each assigned to a different $f$ valency  indicate that there is more than just a single component. The $f$ configurations are mixed in the ground state, but the degeneracy is lifted in the XPS or XAS final state, as the core-hole is screened differently by the $\left|f^0\right>$, $\left|f^1\right>$, or $\left|f^2\right>$ states. This should normally allow a direct estimation of the various $f$-electron weights, and therefore help to characterize the degree of hybridization of the $f$-electron, at the origin of the heavy-fermion like behavior of Ce.  However, the electron-count derived in the final state is not representative of the Ce ground state and the Ce electronic properties at the $\gamma$-$\alpha$\ transition have, in fact, never been observed \emph {directly} in the bulk material.

Recently, resonant inelastic x-ray scattering (RIXS) has emerged as a means of probing  the mixed-valent behavior in rare-earths in considerable detail~\cite{Dallera2002}.  The experiment consists in measuring the $3d \rightarrow 2p$ decay following a resonant excitation close to the rare-earth $L_{2,3}$ edge ($2p \rightarrow 5d$ transition). The RIXS process, subsequently denoted $2p3d$-RIXS, benefits from the selective resonant enhancement of the different valent states, through a proper choice of the incident energies.  In a previous series of experiments, we have studied the $\gamma$-$\alpha$\ transition with temperature in Ce-Sc and Ce-Th alloys by $2p3d$-RIXS~\cite{Rueff2004a}.  The chemical pressure induced by alloying normally prevents contamination by the intermediate $\beta$-phase. The sudden changes in the electronic properties at the transition temperature and the accompanying hysteresis both pointed to a first-order transition, consistent with a pure $\gamma$-$\alpha$\ isostructural change. However, electron interactions with the dopant element (Sc or Th) necessarily intervenes in the Ce-$4f$ electronic properties.  

Here we investigate elemental Ce by $2p3d$-RIXS directly subjected to high pressure to induce the $\gamma$-$\alpha$\ transition.  Without the alloying effects inherent in previous experiments~\cite{Rueff2004a,Dallera2002} it has now become possible to implement  the Anderson impurity model (AIM) to analyze the spectra and, from there, estimate the ground state $f$-counts  in both phases and in particular the variation in $n_f$ and double occupancy across the transition. This combination of RIXS spectroscopic measurements with advanced simulations provides the most complete picture of the bulk electronic properties of elemental Ce at the $\gamma$-$\alpha$\ transition to date.

The starting material was an ingot of high-purity polycrystalline Ce kept in silicon oil.  Maximum care was taken to avoid oxidation of the sample during the loading and measurement phases.  A small chip of Ce was cut off directly in silicon oil and loaded in a membrane diamond anvil cell with rubies for pressure calibration. The oil served both as pressure transmitting medium and to prevent oxidation of the sample  once in the pressure cell.  At the energies of the Ce $L_3$ edge (5.73~keV) and of the $3d \rightarrow 2p_{3/2}$ ($L\alpha_1$) emission (4.84~keV), absorption of x-ray by diamonds is critical.  The pressure cell was equipped with a pair of perforated diamonds each caped with a 0.5~mm thick diamond to maximize throughput. We used Rh as gasketting material (for details, see Ref.~\cite{Itie2005}).
The RIXS experiments were performed at ID-12 at ESRF. The incident energy was selected by a fixed-exit Si(111) monochromator and the beam focused at the sample position in a 40$\times$300 (V$\times$H)~$\mu\mathrm{m}^2$ spot.  We privileged the transmission geometry with $\approx 10^{\circ}$ scattering angle.  This configuration optimizes the scattering volume and hence the emitted signal strength. It also ensures that  the sample probed is homogeneous with respect to pressure gradient.  The RIXS spectra were acquired using a bent crystal spectrometer~\cite{Hague2000}. The pressure cell was mounted outside the spectrometer, with the emitted x-rays entering the spectrometer chamber through a thin Be window. Count rates at the maximum of the Ce $L\alpha_1$ line were of the order of 10~Hz because of the low transmission by the pressure cell.

\begin{figure}[htb]
\includegraphics[width=8.5 cm]{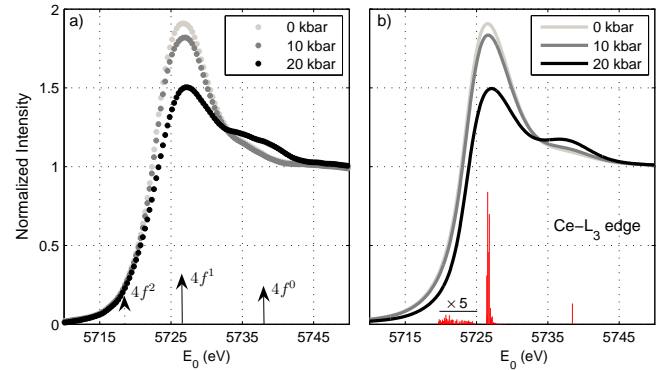}
\caption{(color online). Experimental (a) and calculated (b) $L_3$ XAS spectra in elemental Ce as a function of pressure. The spectra are normalized to a unity jump at the absorption edge.  Solid arrows point to the $4f^0$ and $4f^1$ components.  The dashed arrow indicates the incident energy where the $4f^2$ component resonates in the $2p3d$-RIXS spectra. Ticks in panel b) are the multiplet states (shown at 0 kbar).}
\label{fig:XAS}
\end{figure}

Figure~\ref{fig:XAS}a shows the experimental $L_3$ XAS spectra as a function of pressure.  
The white line shows a marked decrease in intensity as Ce is driven through the $\gamma$-$\alpha$ transition, while the feature denoted $4f^0$ progressively builds up at higher energy.  The overall spectral shape and the spectral changes at the transition are consistent with early results by Lengeler \emph{et al.}~\cite{Lengeler1983}.  In the latter, however, the $\alpha$-phase was obtained from $\gamma$-Ce by  low temperature quenching and the presence of parasitic $\beta$-phase cannot be ruled out.  Following the interpretation from Gunnarsson and Sch\"onhammer~\cite{Gunnarsson1985}, the XAS final states split into multiple components because of the $2p$ core-hole Coulomb potential acting on the mixed-valent ground state $c_0 |4f^0\rangle + c_1 |4f^1\rangle + c_2 |4f^2\rangle$, where  $|c_n|^2$ represent the weight of the individual components and the superscripts indicate the dominant configuration in each case.  As an example, the white line is mostly attributed to the $4f^1$ final-state configuration, while the high-energy hump is mainly $4f^0$-like.  We use this standard shorthand notation for the subsequent referencing of the spectroscopic features.  Note that the $4f^2$ component, expected to show up in the pre-edge region, is masked by the $2p_{3/2}$ core-hole lifetime.

\begin{figure}[htb]
\includegraphics[width=8 cm]{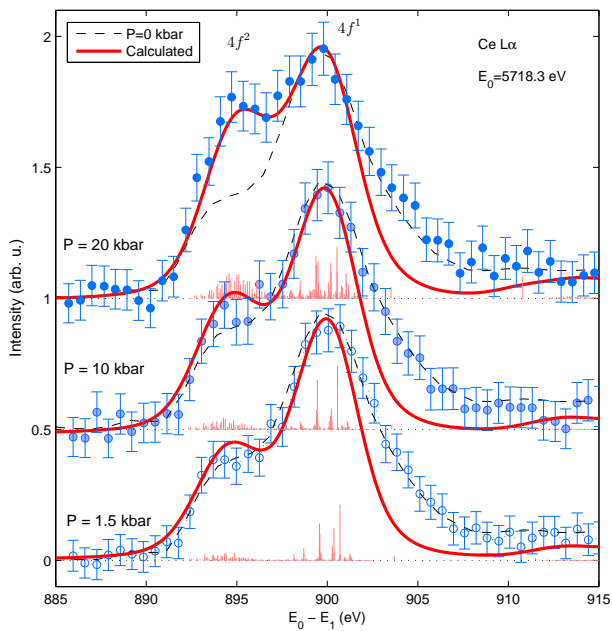}
\caption{(color online). $2p3d$-RIXS spectra in elemental Ce as a function of pressure (circles).  All the spectra were measured at a fixed incident energy $E_0$ set to 5718.3~eV.  To emphasize changes with pressure, the spectrum measured at ambient conditions (dashed line) is repeated for each pressure curve. Superimposed are the calculated spectra (thick lines).  The spectra are normalized to the maximum intensity and offset for clarity. Ticks indicate the multiplet states.}
\label{fig:RIXS}
\end{figure}

A series of $2p3d$-RIXS spectra were measured at 0~kbar in the pressure cell at incident energies $E_0$ in the Ce $L_3$ edge region.  The general behavior of the RIXS spectra as a function of the incident energy follows closely the results obtained in Ce-Sc and Ce-Th alloys in the  $\gamma$-phase~\cite{Rueff2004a,Dallera2004a}. The series is not reproduced here, but we briefly summarize the main findings: as the incident energy is tuned to the pre-edge region, a shoulder shows up in the RIXS spectra on the low energy-transfer side of the main emission (see Fig.~\ref{fig:RIXS}); this feature resonates at $E_0=5718.3\;{\mathrm{eV}}$ (denoted $4f^2$ in Fig.~\ref{fig:XAS}); at higher incident energies, the main line continues growing in intensity until it reaches a maximum at the white line energy ($4f^1$ label in Fig.~\ref{fig:XAS}), where the fluorescence regime sets in.  
Accordingly the low-energy shoulder observed in the RIXS spectra at 5718.3~eV can be traced back to a $4f^2$-dominated final state (see Fig.~\ref{fig:RIXS}), while the main emission line comes from transitions to mainly $4f^1$ final states.  In Fig.~\ref{fig:RIXS}, we can follow the evolution of the $2p3d$-RIXS spectra measured at $E_0=5718.3\;{\mathrm{eV}}$ as the pressure is increased.  The spectrum at 1.5~kbar barely shows a difference with the standard pressure data.  However, a striking increase ($\approx 40$\%) in the $4f^2/4f^1$ intensity ratio is observed as the systems passes the $\gamma$-$\alpha$ transition pressure.

\begin{table*}[htb]
\caption{Parameters used in the calculations as a function of pressure and volume (after Ref.~\cite{Jeong2004}): $\epsilon_f^0$ is the binding energy of the bare $f$-level and $V$ the hybridization energy. Weights of the $4f$-components and the resulting occupation number $n_f$, and variation $\Delta n_f$ are also indicated. $T_{\mathrm{K}}$ is the Kondo temperature.}
\label{tab:Parameters}
\begin{ruledtabular}
\begin{tabular}{ccccccccccccc}
$P$ (kbar) & $v$ (\AA$^3$) & $\epsilon_f^0$ (eV) & $V$ (eV) & $4f^0$ (\%) & $4f^1$ (\%) & $4f^2$ (\%) & $n_f$ & \multicolumn{4}{c}{$\Delta n_f$ (\%)} & $T_{\mathrm{K}}$ (K)\\
\hline
 0 & 34.4 & $-1.3$ & 0.31 & 5.1 & 92.5 & 2.4 & 0.97 & 0\footnotemark[1] & 0\footnotemark[2] & 0\footnotemark[3] & 1.5\footnotemark[4]& 70$\pm$10 \\
10 & 27.6 & $-1.2$ & 0.34 & 7.7 & 89.6 & 2.7 & 0.95 & $-2$\footnotemark[1] & $-4.5$\footnotemark[2] & $-10$\footnotemark[3] & 8\footnotemark[4]& 200$\pm$50 \\
20 & 26.7 & $-0.8$ & 0.43 & 21.9 & 74.7 & 3.4 & 0.81 & $-17$\footnotemark[1] & $-5.5$\footnotemark[2] & - & 10\footnotemark[4]& 1700$\pm$200 \\
\end{tabular}
\end{ruledtabular}
\footnotetext[1]{This work.}
\footnotetext[2]{DMFT at 632 K, from Ref.~\onlinecite{McMahan2003}}
\footnotetext[3]{DMFT at 400 K, from Ref.~\onlinecite{Zolfl2001}}
\footnotetext[4]{DMFT at 400 K, from Ref.~\onlinecite{Amadon2005}}
\end{table*}
In a previous experiment, we had extracted the corrected $4f^2/4f^1$ ratio after deconvolution from lifetime broadening using a phenomenological approach. Here, we take a major step forward  by carrying out full multiplet calculations within the Anderson impurity model. 
Details of the model calculations and Hamiltonian have been described in previous work~\cite{Kotani2001,Ogasawara2000a}. We use a basis set consisting of $4f^0$, $4f^1\underline{v}$, and $4f^2\underline{v}2$ configurations in the ground state, where $\underline{v}$ denotes a hole in the valence band below the Fermi energy ($\varepsilon_F$). The intermediate states are thus described by linear combinations of $2p^54f^05d$, $2p^54f^1\underline{v}5d$, and $2p^54f^2\underline{v}^25d$. The final state contains $3d^94f^05d$, $3d^94f^1\underline{v}5d$, and $3d^94f^2\underline{v}^25d$.
The atomic Slater integrals and spin-orbit interaction parameters are obtained using Cowan's Hartree--Fock program with relativistic corrections~\cite{Cowan1981} and scaled to 80\% to account for intra-atomic configuration interaction effects. 
The configuration-dependent hybridization strength with ($R_c$,$R_v)= (0.6,0.8)$ are also used in the present analysis, following Ref.~\cite{Okada1995}. The parameters of $U_{ff}$ (on-site Coulomb repulsion), $U_{fc}(2p)$ (($U_{fc}(2p)=U_{fc}(3d)+0.5$ eV)  is the attractive core-hole potential), and $W$ (conduction-band width) are taken as 6.0, 10.5 eV, and 2.0 eV respectively. The hybridization strength $V(\varepsilon)$ between the $4f$ and the conduction band (CB) states depends on the CB energy $\varepsilon$. The CB states are supposed symmetric around $\varepsilon_F=0$ and we use a semi-elliptical form of $V(\varepsilon)^2$ discretized on a logarithmic $\varepsilon$-scale~\cite{Nakazawa2002}. The RIXS cross-section is derived from the Kramers-Heisenberg formula (see Eq.~5 in Ref.~\cite{Kotani2001}). 

The simulated XAS and RIXS spectra are represented in Fig.~\ref{fig:XAS}b and Fig.~\ref{fig:RIXS} respectively for the three measured pressures.  The XAS spectra are well reproduced throughout the transition. The overall agreement for RIXS is equally good, except on the high energy-transfer side.  The discrepancy likely results from a fluorescence-like contribution to the spectra, which is not taken into account in the calculations. The parameters used in the calculations are reported in Table~\ref{tab:Parameters} for pressures of 0~kbar, 10~kbar, and 20~kbar, along with the estimated weights for the different $4f^n$-components. The parameter self-consistency was checked against XAS, RIXS spectra, and valence-band photoemission data from Ref.~\cite{Kucherenko2002}. These were shown to be already well described by a simplified version of IAM, which nevertheless takes into account the actual density of states~\cite{Hayn2001}.

The main effect, according to the calculation, is the sharp decrease in the $4f^1$ component to the advantage of the $4f^0$-related feature, which gains intensity as Ce becomes more $\alpha$-like.  Such a trend is consistent with the spectral changes in the XAS spectra.  Formally, the transfer of spectral weight from the $4f^1 (5d^1)$ configuration toward a more $4f^0 (5d^2)$ configuration in the $\alpha$-phase can be understood as a partial delocalization of the $4f$ electrons. Interestingly enough, the highly hybridized $4f^2$ state also shows a sizable ($\sim$40\%) increase with pressure. The increased contribution from the $4f^2$ component at high pressure stresses the growing interaction between the $4f$ and the conduction electrons, a characteristic feature of Kondo-like behavior. This is supported by the increase of the hybridization parameter $V$ as reported in Table~\ref{tab:Parameters} in the high pressure regime. The growth of double occupancy at low volume has another important consequence: it points to less correlation in the $\alpha$-phase as electron hopping is favored. Therefore, the new picture that arises from the RIXS analysis at the $\gamma$-$\alpha$  transition is that of the coexistence of competing effects: partial delocalization of the $4f$ electrons through band formation with the conduction states on the one hand, and reduced electron-electron correlations on the other hand that allows the system to accommodate stronger on-site repulsion. 

Finally, the change  in $n_f$ is obtained from the calculated weights of the $4f$-components, according to $n_f = 0 |c_0|^2 + 1 |c_1|^2 + 2 |c_2|^2$.  We obtain $n_f = 0.97$ in the $\gamma$-phase ($P=1.5\;\mathrm{kbar}$), and $n_f = 0.81$ in the $\alpha$-phase ($P=20\;\mathrm{kbar}$). The results are clearly consistent with earlier estimations~\cite{Wuilloud1983,Liu1992,Dallera2004a} for the $\gamma$-phase, but not for the $\alpha$-phase where values differ substantially. Our RIXS-derived $n_f$-value is 10--15\% lower. These new values of the $f$-occupation for Ce are particularly opportune  as they can be compared to recent {\textit{ab initio}} calculations.  Table~\ref{tab:Parameters} compares the change in $n_f$ at the transition derived from the RIXS data to values calculated within dynamical mean-field theory (DMFT) as a function of volume change. (We use the equation of state derived by Jeong \emph{et al.}~\cite{Jeong2004} in Ce for pressure to volume conversion.)  The discontinuous dependence of $n_f$ at the transition is well accounted for by DMFT~\cite{McMahan2003,Zolfl2001} in the low temperature limit.  On the other hand, the drop in $n_f$ at the transition is largely underestimated (4--10\% in the DMFT calculations against $\approx 20$\% according to the RIXS results).  The discrepancy with our  experimental finding calls for an improved basis set and functional.   

$n_f$ deviates from unity in Ce as a direct consequence of non-zero hybridization.  A remarkable manifestation of this Kondo behavior is the occurrence of a quasiparticle resonance at $E_F$ in the single-particle spectral function $\rho_f(\omega)$.  According to the Friedel sum rule, $\rho_f(E_F)$ depends only on the ground state value of $n_f$, the degeneracy $N_f$ of the local orbital, and a coupling parameter $\Delta \propto V^2/W$, and varies as~$n_f(1-n_f)$.  The sharp decrease of $n_f$ in the $\alpha$-phase indicates a strong enhancement of the quasiparticle peak and that of the renormalization of the bare particle corresponding to $(1-n_f)$.  
But the former effect is partly smeared out at temperatures comparable to the Kondo temperature $T_{\mathrm{K}}$~\cite{McMahan2003}.  $T_{\mathrm{K}}$ is here the key quantity to characterize the $4f$-electron coupling with the Fermi sea.  It can be evaluated thanks to the Friedel sum rule and given the approximate relationship $(1-n_f)/{n_f}\sim(\pi k_B T_K)/(N_f\Delta)$~\cite{Gunnarsson1983} in the limit of large $N_f$.  We obtain 70~K in the $\gamma$-phase and 1700 K in the $\alpha$-phase assuming $\Delta\sim110$~meV.  The temperatures show a fair agreement with neutron scattering data~\cite{Murani1993} obtained in Ce-Sc alloys but differ very significantly from the generally accepted XPS-derived values~\cite{Liu1992}. They are consistently smaller by a factor $\sim 2$ in the $\alpha$-phase. 
\begin{acknowledgments}
We thank B.~Amadon for fruitful discussions on DMFT calculations.
\end{acknowledgments}

\end{document}